\documentclass[fleqn]{article}
\usepackage{espcrc2}
\usepackage{psfig}
\newcommand{\BR}{{\cal B}}

\newcommand{\jpsi}{J/\psi}
\newcommand{\psp}{\psi(2S)}
\newcommand{\pspp}{\psi(3770)}
\newcommand{\psppp}{X(3872)}
\newcommand{\psippp}{X(3872)}

\newcommand{\chico}{\chi_{c1}}

\newcommand{\jpsipp}{\pi^+\pi^-J/\psi}
\newcommand{\ppjpsi}{\pi^+\pi^-J/\psi}
\newcommand{\EE}{e^+e^-}
\newcommand{\MM}{\mu^+\mu^-}

\newcommand{\pp}{\pi^+\pi^-}
\newcommand{\ppb}{p\overline{p}}
\newcommand{\LL}{\ell^+\ell^-}

\newcommand{\ra}{\rightarrow}

\newcommand{\gee}{\Gamma_{e^+e^-}}
\newcommand{\gtot}{\Gamma_{tot}}
\newcommand{\Bgee}{\Gamma_{e^+e^-}\BR_{\ppjpsi}}
\newcommand{\beq}{\begin{equation}}
\newcommand{\eeq}{\end{equation}}

\newcommand{\beqn}{\begin{eqnarray}}
\newcommand{\eeqn}{\end{eqnarray}}
\newcommand{\beqns}{\begin{eqnarray*}}
\newcommand{\eeqns}{\end{eqnarray*}}
\newcommand{\bfg}{\begin{figure}}
\newcommand{\efg}{\end{figure}}
\newcommand{\bitm}{\begin{itemize}}
\newcommand{\eitm}{\end{itemize}}
\newcommand{\bnum}{\begin{enumerate}}
\newcommand{\enum}{\end{enumerate}}
\newcommand{\btbl}{\begin{table}}
\newcommand{\etbl}{\end{table}}
\newcommand{\btbu}{\begin{tabular}}
\newcommand{\etbu}{\end{tabular}}

\title{The upper limit of the $\EE$ partial width of $\psppp$}
\author{C.~Z.~Yuan \address[IHEP]{Institute of High Energy Physics,
P.O.Box 918, Beijing 100039, China}
\thanks{Supported by 100 Talents Program of CAS (U-25)},
X.~H.~Mo \addressmark[IHEP]$^,$\address[CCAST]{China Center of Advanced
Science and Technology, Beijing 100080, China} , 
P.~Wang \addressmark[IHEP]
}

\date{\today}
\begin{document}

\begin{abstract}

The $\EE$ decay partial width of the recently observed state, 
$\psippp$, is estimated using the ISR data collected at
$\sqrt{s}=4.03$~GeV in $\EE$ annihilation experiment by BES at
BEPC. It is found that $\Bgee<10$~eV at 90\% C. L. if the
$J^{PC}$ of $\psippp$ is $1^{--}$. Together with the potential
models and other information, we conclude that $\psippp$ is very
unlikely to be a vector state.

\end{abstract}

\maketitle

\section{Introduction}

Belle recently reported a new state at 3872~MeV (denoted as $\psippp$) in
$\ppjpsi$ invariant mass spectrum in $B^{\pm}\ra K^{\pm}
(\ppjpsi)$, besides the huge signal of $\psp$ at
3686~MeV~\cite{Belle}. This was soon confirmed by CDF in the
inclusive mass spectrum of $\ppjpsi$ in $\ppb$ experiment at
Tevatron~\cite{CDF}.

The small width and the mass very close to the $D \overline{D^*}$
mass threshold are of great interest and there have been various
interpretations of this state, as the $1^3D_2$ state of the
charmonium, as the $D \overline{D^*}$ molecular, as the mixture of the $1^3D_2$ charmonium and $D \overline{D^*}$ molecular, as the $h_c^\prime(^1P_1)$, 
as the diquark-diquark bound state, or the deuson and so
on~\cite{voloshin,pakvasa,tornqvist,close}. Among these possible
interpretations, the $1^3D_2$ state of the charmonium state
has gained great weight due to its naturalness, and coincidence with
the potential model prediction, and its forbidden decay to $D
\overline{D}$ due to parity conservation. The CDF result on
production rates of this state and $\psp$ in $\ppb$ experiment
also supports $\psippp$ being a natural state~\cite{ktchao}.
However, this interpretation will result in big decay branching
fraction to $\gamma \chico$, which was found to be in contradiction with
the Belle measurement~\cite{Belle}.

The possibility of $\psppp$ being a vector charmonium
state is believed to be faint because the typical
width of a vector charmonium state at this mass is around
a few ten MeV and its decays to charmed mesons will be dominant.
However, there is no direct experimental test on this
hypothesis. It has also been suggested in Ref.~\cite{close}
that BES or CLEOc search for this state in $\EE$ annihilation
in the vicinity of its mass to rule out this possibility
(or very unlikely to establish its $J^{PC}$ as $1^{--}$).
While high precision experimental information from BES and
CLEOc will certainly improve the situation greatly, the 
existing experimental
result in literature has already given us some information on
this, that is, the Initial State Radiation (ISR) events 
collected at higher energy experiments.

It is of great interest to note that using 22.3~pb$^{-1}$ data at
$\sqrt{s}=4.03$~GeV from BES, through $\ppjpsi$ events with $\jpsi$
decays into lepton pairs, an extensive study was
made~\cite{BES4.03}, which includes the searching for the possible
metastable hybrids ($q \overline{q} g $) produced in $\EE$
annihilation, the searching for the massive charmonium state
$\psi(3836)$, the measuring of the $\EE$ partial width of $\psp$, 
and so forth. If $\psippp$ is a $1^{--}$ state, 
it can be produced in the same final states in this data sample with even 
larger effective luminosity comparing with $\psp$,
since $\psppp$ is closer than $\psp$ to the center of mass energy
$4.03$~GeV.

In this Letter, the number of detected $\psippp\ra \jpsipp$ events
$n^{obs}$ is obtained from the ISR data at
$\sqrt{s}=4.03$~GeV. The production cross section $\sigma^{prod}$ is
evaluated theoretically taking into account the ISR and
the energy spread of the experiment. Using above two numbers, 
the upper limit of the $\EE$ partial width of $\psippp$ is obtained with
the estimation of the branching fraction of $\psippp \ra \jpsipp$. At last, 
possible ways to further refine the result to have a better understanding
of the nature of $\psippp$ state are suggested.

\section{Evaluation of the observed number from ISR data}

\begin{figure}[htbp]
\centerline{\hbox{ \psfig{file=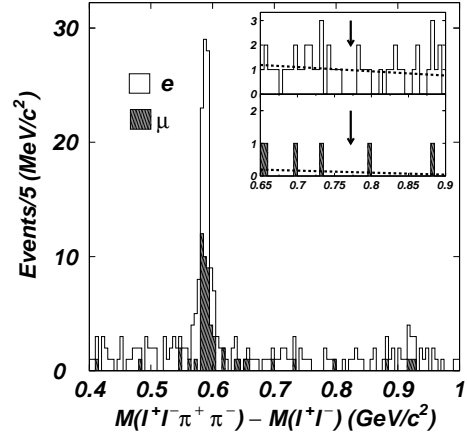,width=6.0cm}}}
\caption{The invariant mass difference between $\pp \LL$ and $\LL$
from BES experiment (Figure 1 of Ref.~\cite{BES4.03}). The blank
histogram is for $\EE$ and the hatched one for $\MM$. The insets
show the fits of the plots in $\psippp$ mass region, the
dotted lines are from the best fits, and the arrows show the
position of $\psippp$ state.} \label{datfit}
\end{figure}

Using ISR data collected by BES detector~\cite{bes}, the final
state $\jpsipp$ was studied, where $\jpsi$ resonance is
tagged by lepton pairs, either $\EE$ or $\MM$~\cite{BES4.03}. 
A $\jpsi$ candidate,
defined as the dilepton invariant mass between 2.5 and 3.25~GeV,
is combined with a pair of oppositely charged tracks, where at
least one track should be identified as a pion according to the
energy loss ($dE/dx$) in the main drift chamber and the time-of-flight 
measurements. The difference in invariant mass between $\pp \LL$ and $\LL$ 
($\ell = e,\mu$) is shown in Fig.~\ref{datfit} (reproduced from Fig. 1 of
Ref.~\cite{BES4.03}) for the two decay modes. The prominent peaks around 0.6 
GeV correspond to $\psp \ra \ppjpsi$, $\jpsi \ra \EE$ and $\MM$ decays.

For the resonance $\psippp$, which corresponds to a mass
difference from $\jpsi$ of 0.775~GeV, there is no signal in either
$\EE$ or $\MM$ channel, as can be seen in Fig.~\ref{datfit} (the
insets are the details of the figure). In the following, we will
try to determine the upper limits of the numbers of $\psippp$
events.

Our fit is performed for both $\EE$ and $\MM$ modes for the mass
differences ranging from 0.65 to 0.9~GeV, with a linear
background and a Gaussian smeared Breit-Wigner (BW) for the signal
using maximum likelihood method. In the fitting, the resonance
mass is fixed at 3.872~GeV according to Ref. \cite{Belle}, and the
mass resolution is set to be 9.4~MeV by the measurement at $\psp$
in Ref.~\cite{BES4.03}. So far as the total decay width
$\Gamma_{tot}$ is concerned, Belle gave a BW width parameter
$\Gamma_{tot}= (1.4 \pm 0.7)$~MeV, from which, the upper limit of
$\Gamma_{tot}< 2.3$~MeV was inferred at 90\% confidence level
(C.~L.). In our study, the values $\Gamma_{tot}= 2.3$~MeV (the
upper limit of the Belle measurement), $\Gamma_{tot}= 1.4$~MeV
(the central value of the Belle measurement) and $\Gamma_{tot}=
0.23$~MeV (the typical width of non-$D \overline{D}$ decay charmonium
states), are attempted in the following evaluations.

With these parameters, the fits yield nought signal events in both
$\EE$ and $\MM$ channels, almost independent on the $\Gamma_{tot}$
used. The upper limits of the numbers of the observed events from
$\psippp$ decays at 90\% C.~L. are listed in Table~\ref{evalnmb}.

\begin{table}[htb]
\caption{\label{evalnmb} Upper limits of the numbers of the observed
events from $\psippp$ decays at 90\% C.~L.}
\begin{tabular}{lcc} \hline \hline
    $\Gamma_{tot}$ (MeV)  & $\EE$  Mode  & $\MM$  Mode  \\\hline
                    2.3   & 5.98         & 1.92         \\
                    1.4   & 5.91         & 1.90         \\
                    0.23  & 5.81         & 1.86         \\ \hline \hline
\end{tabular}
\end{table}

From Table~\ref{evalnmb}, it can be seen that the effect due to
different $\gtot$ is rather small. As a conservative estimation,
the largest numbers are used as the upper limits for the
numbers of the observed events, that is, at 90\% C.~L.,
\beq
 n^{obs}_{\pp\EE} < n^{up}_{\EE}= 5.98 ~,
\label{upee}
\eeq
and
\beq
 n^{obs}_{\pp\MM} < n^{up}_{\MM}= 1.92 ~.
\label{upuu}
\eeq

\section{Theoretical calculation of the production cross section}

In $\EE$ annihilation experiment at the center of mass energy $\sqrt{s}$, 
the cross section of resonance $\psippp$ at the Born order is expressed by 
the BW formula
\begin{equation}
\sigma^{Born}(s)=\frac{12\pi \gee \gtot}{(s-M^{2})^{2}
+\gtot^2 M^{2}}~,
\label{BW}
\end{equation}
where $M$ and $\gtot$ are the resonance mass and the total width
of $\psippp$ respectively, and $\gee$ is the partial width of 
$\psippp \ra \EE$. 

The production cross section of $\psippp$ due to ISR from experiment
operating at the center of mass energy $\sqrt{s_0}$ can be
calculated by
\begin{equation}
\sigma^{prod} (s_0) =\int \limits_{x_{low}}^{x_{up}} dx
F(x,s_0) \frac{\sigma^{Born}(s_0(1-x))}{|1-\Pi (s_0(1-x))|^2}~,
\label{RAD}
\end{equation}
where $F(x,s_0)$ has been calculated to an accuracy of
0.1\%~\cite{rad.1,rad.2,rad.3},  $\Pi(s)$ is the vacuum
polarization factor~\cite{vacuum}, 
$x_{up}$ and $x_{low}$ denote the superior
and inferior limits of the integration, which are defined as
$$ x_{up} = 1 - \frac{s_{low}}{s_0}~,$$
and
$$ x_{low} = 1 - \frac{s_{up}}{s_0}~.$$
$s_{up}$ and $s_{low}$ correspond to the fitting range of the
experimental data in Fig.~\ref{datfit}, that is
$$ \sqrt{s_{up}} - M_{\jpsi} = 0.9 \hbox{ GeV}~,$$
and
$$ \sqrt{s_{low}} - M_{\jpsi} = 0.65 \hbox{ GeV}~,$$
where $M_{\jpsi}$ is the $\jpsi$ resonance mass. 
In unit of keV,
\beq \gee = \alpha \cdot 1 \mbox{ keV.}
\label{geedef} \eeq
Fix the mass and total width to the values
used above (from Belle~\cite{Belle}), the integration gives the
production cross section
\beq
 \sigma^{prod} ( [4.03 \hbox{GeV}]^2)=
      \alpha \cdot 0.61 ~\hbox{nb}~.
\label{sigprod} \eeq 
It should be pointed out that varying 
$\gtot$ has little effect on $\sigma^{prod}$, the integration
with different $\gtot$ actually gives the same value up to
the significant digits listed in Eq.~(\ref{sigprod}).

The energy spread effect on cross section is also taken into
account. In fact, the energy spread hardly affects the calculated
cross section, because the energy spectrum of the ISR photon is
already very flat in the expected $\psippp$ mass region. For
example, if the energy spread is 1.5~MeV at 4.03~GeV~\cite{beamspread},
the difference between the cross sections with and without energy
spread is at the level of $10^{-4}$ relatively. So the production cross
section given in Eq.~(\ref{sigprod}) without energy spread
correction, is accurate enough for our following estimations.

\section{Estimation of the $\EE$ partial width}

If the number of the produced $\psippp$ events is denoted as
$n^{prod}$, and the final state $\pp\LL$ is used in the
experiment observation, the relation between $n^{prod}$ and
$n^{obs}_{\pp \LL}$ is expressed as 
\beqn
\lefteqn{ n^{obs}_{\pp \LL} = n^{prod} \cdot } \nonumber \\
&&{\cal B}^{\psippp}_{\pp \jpsi} \cdot {\cal B}^{\jpsi}_{\LL}
  \cdot \varepsilon_{\pp \LL}~,
\label{obsprod} \eeqn
where ${\cal B}^{\psippp}_{\pp \jpsi}$  is
the branching fraction of $\psippp \ra \pp \jpsi$, ${\cal
B}^{\jpsi}_{\LL}$ the branching fraction of $\jpsi \ra \LL$, and
$\varepsilon_{\pp \LL}$ the efficiency of detecting 
$\pp \LL$ final state.

$n^{prod}$ can also be expressed by
\beq
 n^{prod} = {\cal L} \cdot \sigma^{prod}~,
\label{lprod} 
\eeq 
with ${\cal L}= 22.3$~pb$^{-1}$, which is the
integrated luminosity of the data taken at 4.03~GeV~\cite{BES4.03},
and $\sigma^{prod}$ is given in Eq.~(\ref{sigprod}).

With Eqs.~(\ref{upee}) and (\ref{upuu}), it is obtained
\beq
\sigma^{prod} \cdot {\cal B}^{\psippp}_{\pp \jpsi} < \frac{
n^{up}_{\LL} } { {\cal L} \cdot {\cal B}^{\jpsi}_{\LL} \cdot
\varepsilon_{\pp \LL} } ~.
\label{lesseq}
\eeq

According to PDG~\cite{pdg}, ${\cal B}^{\jpsi}_{\EE}= (5.93\pm
0.10)\%$ and ${\cal B}^{\jpsi}_{\MM}= (5.88\pm 0.10)\%$. As an
estimation, the efficiency of $\pp \LL$ final state from $\psippp$
decay is treated as the same as that from $\psp$~\footnote{This
should be an underestimation of the efficiency since the momentum
of the pion tracks from $\psippp$ decays will be more energetic
and the detection efficiency will be larger than in $\psp$ case.
This leads to an overestimation of the upper limit of $\gee$, so the
numbers we obtained will be conservative.}:
$\varepsilon_{\pp\EE}=(22.9\pm 0.1)\%$ and
$\varepsilon_{\pp\MM}=(18.9\pm 0.1)\%$~\cite{BES4.03}. 
Then the product of $\gee$
and ${\cal B}^{\psippp}_{\pp \jpsi}$ is acquired
\[ \Bgee^{\psippp}< 30~\hbox{eV}~, \]
for $\pp \EE$ final state, and
\[ \Bgee^{\psippp}< 10~\hbox{eV}~, \]
for $\pp \MM$ final state at 90\% C.L.

If we assume that $\Gamma_{\jpsipp}$ of $\psippp$ is about the
same as that of $\psp$ (85.4~keV~\cite{psipscan}), then
$$ {\cal B}^{\psippp}_{\pp \jpsi} >
\frac{85.4 \hbox{ keV} }{2.3 \hbox{ MeV} } = 3.7~\%~,$$ 
so $\gee< 0.82 \hbox{ keV}$ or $\gee < 0.28 \hbox{ keV}$ for $\EE$
or $\MM$ mode, respectively.

Taking the more stringent ones as the final results, we get
\[ \Bgee< 10 \hbox{ eV at 90\% C.~L.}~, \]
and
\[\gee < 0.28 \hbox{ keV at 90\% C.~L.}~, \]
for $\psippp$ state.

\section{Discussion}

A charmonium state with quantum number $J^{PC}=1^{--}$ is either a
$^3S_1$ or a $^3D_1$ state.

In charmonium family, $\jpsi$ and $\psp$ are well established as
$1^3S_1$ and $2^3S_1$ states. If $\psippp$ is a $^3S_1$
state, the only place to be filled into is $3^3S_1$. But there are
some arguments against this assignment: firstly, there is a
relation between the $\EE$ decay partial widths of the $^3S_1$
states of $\psi$ and $\Upsilon$, that is $\Gamma_{ee}(\psi,
n^3S_1) \approx 4 \Gamma_{ee}(\Upsilon, n^3S_1)$, which holds at least
for $n=1$ and $n=2$. Extrapolate this relation to $n=3$, and use
$\Gamma_{ee}(\Upsilon, 3^3S_1)$ from PDG~\cite{pdg}, it is
expected $\Gamma_{ee}(\psi, 3^3S_1) \approx 1.8$~keV. This
contradicts with the upper limit of $\Gamma_{ee}(\psippp)<0.28
\hbox{ keV}$; secondly, $m_{\psi(2^3S_1)}-m_{\psi(1^3S_1)} \approx
m_{\Upsilon(2^3S_1)}-m_{\Upsilon(1^3S_1)}$ (it is 589~MeV for
$\psi$ and 563~MeV for $\Upsilon$). If the same spacing between
$\psi$ and $\Upsilon$ states is extrapolated to
$m_{\psi(3^3S_1)}-m_{\psi(1^3S_1)}$, then the mass of $3^3S_1$
state of charmonium is close to 4~GeV, which is usually assigned
to $\psi(4030)$.

If $\psippp$ is $^3D_1$ state, it is known that $\pspp$ is the
$n=1$ candidate with some mixing of $2^3S_1$ state. The $2^3D_1$
state should be weakly coupled to $\EE$, which is in agreement
with the experimental limit of $\psippp$. However, its mass at
3.872~GeV is too low to accommodate with potential model
predictions~\cite{eichten}.

One more important argument against the assignment of $\psippp$
as a vector meson is that $1^{--}$ charmonium state above open charm threshold 
decays into $D\overline{D}$ copiously, which makes its total width around a
few ten MeV, an order of magnitude greater than the upper limit of
the $\psippp$ width.

In conclusion, $\psippp$ is very unlikely to be a vector state of
charmonium.

There are possible experiments which can further check this.
BES or CLEOc can perform fine scan in the vicinity
of the state to set a more stringent upper limit on the
production rate, independent on the $\jpsipp$ decay branching
fraction of $\psippp$; B-factories can study it using ISR events 
with higher luminosities. Furthermore, the state can be searched 
in more decay channels in B decays, while HERA and Tevatron 
experiments may supply more information on the production
mechanism. All these will help to finally establish the 
nature of $\psippp$ state.

\section{Summary}

Using the ISR events from BES data at $\sqrt{s}=4.03$~GeV,
the product of the $\EE$ partial width and $\psippp\ra
\jpsipp$ decay branching fraction is determined to be
\[ \Bgee< 10 \hbox{ eV at 90\% C.~L.}~, \]
for $\psippp$ state if its $J^{PC}=1^{--}$. With a comparison between $\psi$
and $\Upsilon$ families and predictions of potential models, we conclude that
$\psippp$ is very unlikely to be a vector state.

\section*{Acknowledgments}

We would like to thank Prof. S. T. Xue for discussing and supplying
useful reference.


\begin{thebibliography}{**}

\bibitem{Belle} K.~Abe {\em et al.} (Belle Collab.), hep-ex/0308029; 
            S.-K.~Choi {\em et al.} (Belle Collab.), hep-ex/0309032.

\bibitem{CDF} G.~Bauer representing CDF Collaboration, Quarkonium
              production: new results from CDF, Talk at the second 
	      workshop of the Quarkonium Working Group, Fermilab,
	      Sept. 20-22, 2003. 
	      http://www.qwg.to.infn.it/WS-sep03/WS2talks/plenary/bauer.ps.gz.

\bibitem{voloshin} M.~B.~Voloshin, hep-ph/0309307.

\bibitem{pakvasa} S.~Pakvasa and M.~Suzuki, hep-ph/0309294.

\bibitem{tornqvist} N.~A.~T\"ornqvist, hep-ph/0308277.

\bibitem{close} F.~E.~Close and P.~R.~Page, hep-ph/0309253.

\bibitem{ktchao} K.~T.~Chao, Prompt D-wave charmonium production at the 
Tevatron and the X(3872), Talk at the second workshop of the 
         Quarkonium Working Group, Fermilab, Sept. 20-22, 2003. 
         http://www.qwg.to.infn.it/WS-sep03/WS2talks/prod/chao.ppt.

\bibitem{BES4.03} J.~Z.~Bai. {\em et al.} BES Collaboration,
         Phys. Rev. {\bf D57} (1998) 3854.

\bibitem{bes} J.~Z.~Bai. {\em et al.} BES Collaboration, Nucl. Instr. Meth.
              {\bf A344} (1994) 319.

\bibitem{rad.1} E.~A.~Kuraev and V.~S.~Fadin, Yad. Fiz. {\bf 41}
       (1985) 733 [Sov. J. Nucl. Phys. {\bf 41} (1985) 466].

\bibitem{rad.2} G.~Altarelli and G.~Martinelli, CERN {\bf 86-02} (1986) 47;
        O.~Nicrosini and L.~Trentadue, Phys. Lett. {\bf B196} (1987) 551.

\bibitem{rad.3} F.~A.~Berends, G.~Burgers and W.~L.~Neerven,
        Nucl. Phys.~{\bf B297} (1988) 429; {\it ibid.} {\bf 304} (1988) 921.

\bibitem{vacuum}F.~A.~Berends, K.~J.~F.~Gaemers and R.~Gastmans,
  Nucl. Phys. {\bf B57} (1973) 381;\\
         F.~A.~Berends and G.~J.~Komen, Phys. Lett. {\bf B63} (1976) 432.

\bibitem{beamspread}J.~Z.~Bai {\em et al.}, BES Collaboration,
           Phys. Rev. {\bf  D53}  (1996) 20;\\
C.~Z.~Yuan, B.~Y.~Zhang and Q.~Qin, High Energy Phys. Nucl. Phys. 
{\bf 26} (2002) 1201 (in Chinese).

\bibitem{pdg} K.~Hagiwara {\em et al.} (Particle Data Group),
              Phys. Rev. {\bf D66} (2002) 010001.

\bibitem{psipscan} J.~Z.~Bai. {\em et al.} (BES Collab.),
         Phys. Lett. {\bf B550} (2002) 24.

\bibitem{eichten} See, for example: Y.~Ding, T.~Huang and Z.~Chen,
         Phys. Lett. {\bf B196} (1987) 191;\\
	 J.~L.~Richardson, Phys. Lett. {\bf B82} (1979) 272;\\
	 D.~S.~Kulshreshtha, Nuovo Cimento {\bf A87} (1985) 25.
\end{thebibliography}
\end{document}